\documentclass[sigchi]{acmart}
\AtBeginDocument{%
  }

\copyrightyear{2025}
\acmYear{2025}

\acmConference[SER 594 '25]{Topics in HCI}{May, 2025}{Tempe, AZ}
\acmBooktitle{Human Computer Interactions : Accessibility}

\begin{document}

\title{Insight: Enhancing Mobile Accessibility for Blind and Visually Impaired Users with LLMs}

\author{Joshua Owusu Ansah, Anuj Kapoor, Ayush Khanna, Manvika Vinod, Precious Njeck, Shuai Gao}
\email{jowusuan@asu.edu,anujkap@asu.edu, akhann46@asu.edu, mvinod2@asu.edu, pnjeck@asu.edu, shuaigao@asu.edu}
\affiliation{
    \institution{Arizona State University}
    \city{Tempe}
    \state{Arizona}
    \country{USA}
}

\begin{abstract}
This research paper addresses the limitations of current mobile accessibility services like TalkBack, which provide manual gesture-based sequential feedback to BVI users. Motivated by the promise of large language models (LLMs), this paper introduces Insight, an Android accessibility service that provides natural language interaction and real-time summarization of the screen. The paper performs a within-subject experimental study with users to compare Insight and TalkBack on usability factors. Results show Insight reduced mental effort and task time, and was preferred because of its dialogue interface, but users felt the need for interruption management. Results show LLM-based interfaces can significantly improve mobile accessibility, and describe the potential of hybrid solutions combining gesture and dialogue modalities towards more inclusive design.
\end{abstract}


\received{5 May 2025}
\received[revised]{5 May 2025}

\maketitle

\section{Introduction}
Mobile devices have become essential tools in daily life, yet traditional smartphone interfaces, primarily designed for sighted users, often pose significant accessibility challenges for the over 2.2 billion people globally with vision impairment, including 39 million who are blind \cite{WHO2023}. Although current accessibility features such as screen readers exist, interacting with complex mobile interfaces remains difficult for Blind and Visually Impaired (BVI) users. The increasing reliance on smartphones requires more effective accessibility solutions to ensure equitable access and participation in the digital world.  

The emergence and advancements of large language models presents an opportunity to revolutionize accessibility of mobile user interfaces for visually impaired and blind users. A Large language model (LLM) is a deep learning-based AI model trained on vast text datasets to understand and generate human language. They work by using complex neural networks that are typically based on transformer architectures for contextual learning, which enables them to learn patterns and generate text by predicting the likelihood of the next word in the sequence \cite{Chang2024}. LLMs are being rapidly adopted in both industry and academia due to their wide range of applications. Examples include GPT-4, Bert, Llama, Gemini, etc. \cite{IBM2023}. These models are commonly used as chatbots, virtual assistants, automated content generators, and coding assistants. Crucially, their advanced language capabilities can interpret complex user instructions, respond with contextually relevant feedback, and provide dynamic guidance in real time---surpassing traditional voice-based screen readers. By bridging the gap between human language and UI elements, LLMs empower visually impaired users with more direct, intuitive control of their mobile devices, fundamentally enhancing daily tasks and independence.

Current accessibility tools, such as Google TalkBack, primarily rely on sequential audio feedback and gesture-based navigation. This approach can be slow, requiring users to manually navigate through every screen element. Furthermore, it places a high cognitive load on users, demanding they remember screen layouts and navigate complex interfaces, which can be mentally demanding and lead to task inefficiency. While helpful, these tools often struggle with providing concise summaries or understanding the overall context of a screen, limiting intuitive interaction. Voice assistants can perform actions but lack the ability to summarize screen content or assist directly with complex app navigation. There are alternate interaction methods such as speech-based output, gesture-based navigation, and braille assisted keypads \cite{Kuber2012}; as well as tactile audio and haptic feedback \cite{Wall2006}, etc. However, there are still challenges with interacting with and interpreting certain user interface components such as icons, buttons, and screen regions \cite{Khan2019_blind}.

To address these limitations, we introduce ``Insight,'' an innovative Android accessibility service powered by Large Language Models (LLMs), specifically Google's Gemini 2.0 Flash model. Insight moves beyond sequential feedback by leveraging the contextual understanding and natural language processing capabilities of LLMs. It analyzes screen content, provides context-aware voice summaries, and allows users to interact with their device using natural language commands and questions. Insight aims to perform actions like navigation and typing on the user's behalf while informing them about the screen content using the great summarization capabilities of LLMs \cite{Wang2023_conversational}, creating a more seamless, conversational, and efficient bridge between the BVI user and their mobile device.

\hfill \break

This work is guided by the core research question:

\textbf{RQ:} ``How does an LLM-based Conversational User Interface (CUI) accessibility service compare with conventional screen readers on mobile phones in performing day-to-day tasks, in terms of users' perceived advantages and limitations?''
\hfill \break

Current mobile accessibility tools can be cumbersome for everyday tasks. Investigating whether an LLM-based CUI like Insight can demonstrably improve task efficiency (speed, ease of use) and provide a more intuitive, less cognitively burdensome experience is vital. Positive findings could justify the adoption of LLM-driven solutions, fundamentally enhancing the independence, digital inclusion, and quality of life for millions of BVI users globally.

This research offers several contributions with potential benefits for various stakeholders. Findings suggest that LLM-based conversational UIs like Insight can reduce task completion time and cognitive load, offering enhanced efficiency and ease of use compared to traditional screen readers. This work demonstrates the potential of LLMs as effective screen summarization tools, although the need to account for long-term user familiarity remains. Furthermore, exploring hybrid approaches that combine gesture-based navigation with conversational interaction could offer improved privacy and flexibility. These insights directly benefit: 

\begin{itemize}
    \item \textbf{Accessibility Designers:} By providing evidence for alternative interaction paradigms beyond traditional gestures, informing the design of more intuitive and efficient accessibility solutions. 
    \item \textbf{Mobile Developers:} By highlighting the potential and methods for incorporating advanced, LLM-powered accessibility services into mobile applications and operating systems. 
    \item \textbf{LLM Researchers:} By offering insights into refining LLM capabilities for specific accessibility use cases, particularly in screen summarization and context-aware interaction within mobile environments.
\end{itemize}

\section{Related Work}
\subsection{How Current Accessibility Feature in Mobile Phone Works}

TalkBack (for Android) and VoiceOver (for iOS) are screen readers designed to make smartphones accessible to visually impaired users by providing spoken feedback and touch-based navigation. These features enable users to interact with their devices through gestures such as swiping, tapping, and using multi-finger commands. When enabled, TalkBack and VoiceOver read aloud text, app labels, and notifications, allowing users to navigate interfaces without relying on sight. They also support advanced functionalities like braille displays, voice commands, and customizable verbosity settings to enhance user experience. Additionally, both screen readers integrate with accessibility tools such as magnification and contrast adjustments to improve usability for people with varying levels of vision impairment. While most solutions focus on runtime accessibility support, such as screen readers and assistive services, another crucial aspect is preventing accessibility issues during the design phase of mobile applications. \cite{Zaina2022} emphasize the importance of integrating accessibility-aware UI design patterns early in the development process. Their guidelines provide actionable practices to minimize accessibility barriers from the outset, complementing runtime tools like TalkBack or LLM-based services.

\subsection{Current State of Conversational UI Models}

Conversational user interfaces (CUIs) have emerged as an important means of interaction, allowing users to communicate with digital systems using natural language. Traditional graphical user interfaces (GUIs) require visual engagement, often making them inaccessible to individuals with visual impairments. Assistive technologies like screen readers and voice-over tools have been developed to bridge this gap, but they still present usability challenges. The rise of large language models (LLMs) has introduced new possibilities for enabling conversational interactions that can facilitate a more intuitive and inclusive user experience \cite{Oelen2024}. Any voice based accessibility service can be considered a CUI, hence it is important to consider the state-of-the-art in Conversational User Interfaces while focusing on accessibility.

\subsubsection{Multimodal Large Language Models for UI Navigation}

Recent studies have demonstrated the feasibility of integrating multimodal LLMs into UI navigation tasks. Ferret-UI, for instance, is a multimodal large language model designed for understanding mobile UI screens. It enhances traditional LLM capabilities by incorporating visual and spatial features to perform icon recognition, text identification, and widget interactions \cite{You2024}. These improvements allow the model to process mobile UI elements more effectively, making CUIs more responsive to user needs.

Similarly, \cite{Wang2023_conversational} introduced a method for enabling conversational interaction with mobile UIs using LLMs. Their approach leverages prompting techniques to allow pre-trained LLMs, such as GPT-3, to summarize screens, generate questions, answer UI-related queries, and map natural language instructions to actions. These developments are critical for blind users, as they provide an alternative to traditional screen readers by making UI elements navigable through conversation.

\subsubsection{Natural Language Instruction Mapping for UI Actions}
Another crucial advancement is the ability to map natural language commands to executable UI actions. \cite{Lister2020} developed a model that interprets multi-step voice instructions and translates them into UI action sequences (e.g., navigating to settings, enabling Wi-Fi, or opening applications). This research is particularly relevant for hands-free interaction and accessibility, enabling users to control mobile applications using speech rather than touch-based interactions.

\subsubsection{Screen Summarization and UI Understanding}
Screen summarization is another promising application of LLMs in UI navigation. \cite{Wang2021_screen2words} introduced Screen2Words, a deep learning model that automatically generates concise textual descriptions of mobile UI screens. This model processes multimodal UI data (e.g., images, text, and structural hierarchy) to create summaries, making UI navigation more accessible for visually impaired users.

In addition, \cite{Oelen2024} explored the use of LLMs in scholarly knowledge infrastructure, highlighting their potential to guide users through complex interfaces using natural language assistance. Their work aligns with the broader goal of integrating LLMs into everyday user interfaces to provide real-time assistance and context-aware navigation support.

\subsubsection{Assistive Technology and Accessible Conversational UI}
Accessibility remains a core motivation for developing conversational UI models. \cite{Lister2020} analyzed various accessibility considerations in CUIs for users with disabilities. Their research reviewed design principles for voice assistants, chatbots, and AI-driven CUIs, identifying gaps in cognitive accessibility, screen reader integration, and multimodal interaction. They concluded that LLMs could bridge these gaps by adapting UI interactions to individual user needs in real time.

\cite{Planas2021} further explored model-driven development for multi experience AI-based UIs, emphasizing the need for flexible CUI frameworks that support different interaction modalities (e.g., voice, text, and gesture-based commands). Their research advocates for integrating LLMs within a broader multi experience development platform (MXDP), allowing for seamless transitions between traditional GUI interactions and conversational UI-based interactions.

\subsection{Current State of Accessibility Using LLMs}
Systems like Savant \cite{Ghosh2024, Kodandaram2024} have demonstrated the potential to control applications using spoken commands, eliminating the need to learn intricate shortcuts by using large language models to translate spoken commands, recognize pertinent UI elements, and carry out actions on personal computers. Savant dramatically increased task completion rates and decreased cognitive stress when tested across programs including Google Docs, Excel, and Spotify. According to the study, Savant has the potential to give visually impaired people a consistent, effective, and intuitive digital experience.

GPTVoiceTasker \cite{Vu2024}, for instance, automates complex mobile tasks through voice commands, learning user interactions and dynamically exploring interfaces. This is a powerful capability for users who may have difficulty with the fine-grained control required for multi-step actions. When compared to current approaches, it increases task efficiency by 34.85\% and executes multi-step jobs with an accuracy of 85.7\%. According to an 18-person user study, GPTVoiceTasker received great usability ratings and decreased cognitive burden and task completion time. The study showed that LLM-powered assistive tools can lead to significant improvements in task completion rates and overall efficiency compared to traditional methods, suggesting that LLMs can genuinely enhance the productivity and independence of users with disabilities.

Projects like PROTECT \cite{Costabile2024} demonstrate the potential of conversational AI to improve web accessibility. By allowing users to interact with websites through natural language, these systems can simplify navigation and content consumption. The extension of this conversational approach to mobile devices holds immense promise for mobile accessibility.

For LLM based accessibility in Mobile Apps, Zhe Liu, Unblind Text Inputs: Predicting Hint-text of Text Input in Mobile Apps via LLM developed an LLM-based hint-text generation model called HintDroid, which analyzes the GUI information of input components and user in-context learning to generate the hint-text. HintDroid, an LLM-based system that automatically generates hint-text for text input fields in mobile apps. HintDroid extracts Android view hierarchy to understand UI structure and use GUI-based prompting to help LLMs generate accurate hints. Using in-context and retrieval-based selection functions to increase the accuracy of LLMs in generating the hint-text.

\subsection{Limitations in Existing Research}
Despite these advancements, existing studies primarily focus on desktop-based CUIs, leaving mobile applications largely unexplored \cite{Wang2023_conversational}. While models like Ferret-UI provide improved visual understanding, they rely on structured HTML-based representations, which may not be the most efficient or adaptive approach for real-time mobile interactions \cite{You2024}. 

Screen2Words attempts to incorporate text, images, and UI metadata, further research is needed to optimize real-time processing and accuracy \cite{Wang2021_screen2words}. Their approach is data heavy since more data was involved in the training of their models.

\cite{Lister2020} identified important accessibility design principles, there is limited empirical evaluation of how blind users interact with LLM-powered CUIs in comparison to traditional screen readers.

Despite the significant advances, LLMs still face limitations in the realm of accessibility. While powerful, LLMs can struggle with nuanced contextual understanding and real-time adaptation to dynamic UI changes. This can lead to errors and require user intervention, limiting the overall effectiveness of these tools. The dynamic nature of user interfaces poses a significant challenge. Furthermore, some systems, like earlier versions of Savant, have limitations in handling complex, multi-step commands; it does not ensure universal compatibility or completely remove the need to memorize commands. The research focuses only on Personal Computers. Mobile UIs have different forms of interaction for performing similar tasks, thus the research presents a potential opportunity for similar research in the mobile UIs.

PROTECT was used for web based applications, accessibility on mobile applications is different as the user does not have a keyboard to navigate as well as the screen dimensions are much smaller offering a very different user experience and with it a unique set of challenges.

GPTVoiceTasker is a significant advancement in mobile speech automation, blind users' accessibility issues are not entirely resolved by it, as fundamentally the user needs to see what's displayed on the screen to make a prompt or an automation. Hence this study is not designed for BVI users accessibility. Better screen reader integration designed keeping BVI needs in mind, increased UI resilience, cross-app compatibility, and less dependence on visual components would all be necessary for a more comprehensive solution.

HintDroid looks at providing hints to BVI users and helping with understanding the context of the current UI, but does not act as a complete tool for conversational interaction, navigation and screens summarizations in mobile applications.

The research shows that CUIs can benefit and potentially be applied to interacting with UIs. However, the research wasn't designed for accessibility. Prioritizing accessibility could improve it while using CUI features. Finally, the reliability of LLM-based systems remains a concern. Improved error handling and feedback mechanisms are crucial to ensure that users can effectively recover from errors and maintain control over the system. Robust error handling is essential for building user trust.

\section{Application Design}

Accessibility services in Android are designed to provide alternative or augmented feedback to users with disabilities. This design choice is crucial as it allows the application to \textbf{intercept and analyze the content of the screen} and to \textbf{perform actions on behalf of the user}. Several existing accessibility solutions for BVI users, such as \textbf{TalkBack} and projects like \textbf{BlindSense}, are also implemented as Android accessibility services. This leverages the operating system's built-in mechanisms for interacting with the UI in a non-visual manner. 

\begin{figure}
    \centering
    \includegraphics[width=1\linewidth]{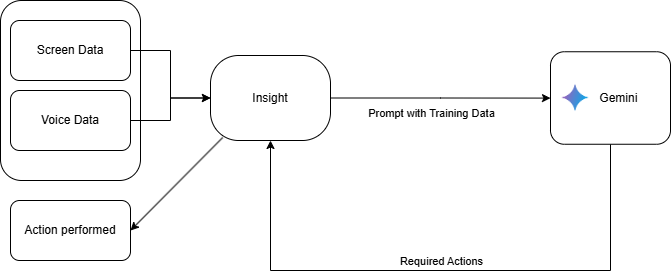}
    \caption{Architecture}
    \label{fig:architectureFig}
\end{figure}

\subsection{Getting the UI tree/JSON data and converting it into a form LLMs can understand.}
The service will utilize the current accessibility features available in Android to collect screen information.
Accessing UI Data with AccessibilityNodeInfo: To access the on-screen data, the service will use Android's built-in class AccessibilityNodeInfo. This class provides information about the nodes in the accessibility hierarchy, as well as the representing UI elements on the screen. 
The information is then converted to a JSON structure and sent to the LLM. JSON was chosen because:
\begin{enumerate}
\item Hierarchical Structure: UI trees inherently have a hierarchical structure (e.g., a screen contains layouts, which contain buttons and text views). JSON is a common and well-understood format for representing hierarchical data using nested objects and arrays.

\item Machine Readability: JSON is easily parsed and processed by machines, making it a suitable format for transferring data between different software components, such as the accessibility service and the LLM.

\item LLM Compatibility: While LLMs are primarily trained on text, they can process structured text formats. Representing the UI tree in JSON allows for a structured and organized way to feed the LLM information about the UI elements, their attributes (like text content, type, position), and their relationships. This structured input can potentially help the LLM better understand the UI.

\item Flexibility: JSON is a flexible format that can accommodate various types of information and can be easily extended if more UI element properties need to be included in the future.
\end{enumerate}

\subsection{Choosing the LLM}
For the project, Gemini 2.0 Flash is chosen. The model is very easily available and has a simple API setup. While models like GPT-4 offer good language understanding, they often come with higher latency and cost, and models like Llama may lack the seamless integration and context caching features Gemini provides. Gemini's tailored API design and lower response times make it suitable for generating conversational, context-aware assistance on mobile devices essential for delivering a smooth, natural user experience for blind and visually impaired users. 

\subsection{Building the prompt for the LLM}

A critical component of the ``Insight'' system is the careful design of the prompt used to guide the Large Language Model (LLM). Prompt engineering is essential for ensuring the LLM behaves predictably, reliably, and safely within the context of an accessibility service interacting directly with the user interface (UI) on behalf of a Blind or Visually Impaired (BVI) user. The prompt defines the LLM's role, inputs, outputs, capabilities, and operational logic. 

\subsubsection{Importance of Structured Interaction}

The primary goal of the prompt is to establish a structured interaction protocol between the Android accessibility service and the LLM. Given the LLM's generative nature, defining strict input and output formats is paramount for system integration \cite{Wen2024, Oelen2024, Liu2024, Wang2023_conversational}. The prompt mandates that the LLM receives the current screen's layout as a JSON representation of the Android accessibility view hierarchy (\verb|screen_context|) and, optionally, the user's transcribed voice command (\verb|user_query|). Crucially, the prompt requires the LLM to \textit{always} respond in a specific JSON format. This structured output, containing fields for \verb|responseType|, spoken \verb|text|, and a list of executable \verb|actions|, allows the accessibility service to unambiguously parse the LLM's intention and execute the necessary steps, whether it's speaking to the user or performing UI manipulations.

\subsubsection{Grounding in Screen Context and Action Specification}

The prompt explicitly grounds the LLM's operations in the provided \verb|screen_context|. This is vital to prevent the LLM from hallucinating UI elements or actions that are not available on the current screen \cite{Liu2024, Wen2024, Wang2023_conversational, Kodandaram2024, Ghosh2024, Vu2024, Song2024}. The prompt details the specific format for the \verb|actions| list, where each action object must specify its \verb|type| corresponding to a predefined set of supported interactions (e.g., ACTION\_CLICK, ACTION\_SCROLL\_FORWARD, ACTION\_SET\_TEXT, open\_app) \cite{Song2024}. These action types directly map to capabilities within the Android Accessibility Service API. For actions targeting specific UI elements, the prompt requires including the element's \verb|bounds| (obtained from the \verb|screen_context|), ensuring precise interaction. Navigation actions are handled distinctly using the \verb|navigationType| parameter. This detailed specification clearly defines the scope of the LLM's ability to interact with the device, ensuring generated actions are valid and executable.

\subsubsection{Defining Interaction Logic and User Experience Considerations}

To handle various interaction scenarios effectively, the prompt defines distinct processing logic based on the inputs received \cite{Wang2023_conversational}:

\begin{enumerate}
    \item \textbf{Screen Context Only:} Trigger a comprehensive screen summary (\verb|responseType: Summarize|).
    \item \textbf{Context + Action Query:} Interpret the query and generate a sequence of UI actions (\verb|responseType: Action|), minimizing spoken output for efficiency.
    \item \textbf{Context + Question Query:} Answer the user's question based on screen content (\verb|responseType: Answer|).
    \item \textbf{Context + Irrelevant/Unprocessable Query:} Generate a helpful error message (\verb|responseType: Error|).
\end{enumerate}
This logic ensures predictable behavior across different use cases \cite{Oelen2024, Lister2020}. Furthermore, the prompt includes general guidelines emphasizing clarity, conciseness, and a conversational tone in the spoken \verb|text|, tailored to the needs of BVI users \cite{Ghosh2024}. It also provides specific instructions for handling ambiguity (e.g., interpreting vague commands as actions) and defines conventions (e.g., mapping scroll directions) to align with user expectations \cite{Song2024}. Robust error handling is mandated, requiring the LLM to explain issues clearly if it cannot fulfill a request (e.g., ``I couldn't find the `Submit' button\ldots'').

The prompt uses a 1-shot technique by providing examples of the situations to the LLM and the expected behavior, for increased accuracy \cite{Liu2024}.

The meticulously crafted prompt serves as the operational blueprint for the LLM within the accessibility service. It ensures contextually grounded, predictable, and user-centric interactions by enforcing structured I/O, clearly defining actionable capabilities, and outlining specific processing logic for various user needs, thereby forming the cornerstone of the Insight system's functionality and reliability.

\subsection{Voice Input and Output}
The native Android speech-to-text implementation captures the user's voice via the device's built-in microphone and converts it to text using a local speech-to-text engine. This text is then sent to the LLM for processing, and the resulting output is converted back into speech by a native text-to-speech engine. The service has a text-only exchange with the LLM to minimize network latency and overcome limitations in upload speed, as transmitting raw audio would introduce significant delays. 

\subsection{Performing the actions}
Once the LLM sends the action list to the accessibility service, the service processes the request and procedurally performs the actions. 
Upon receiving a structured list of actions, the service processes each request sequentially to interact with the device. The screen context, derived from the current state of the user interface, provides a dynamic inventory of possible actions applicable to each interactive element. \cite{Chang2024}
The supported actions within the accessibility service encompasses fundamental UI interactions:
\begin{itemize}
\item Direct manipulation of screen elements, including clicking buttons or selectable items, scrolling through lists or content, typing text into input fields, and selecting text within designated areas. These actions align with the standard interaction modalities observed in mobile operating systems and are essential for navigating and manipulating application content.

\item Navigation directives, such as navigating back through the application history, returning home to the launcher screen, and opening specific applications based on explicit user requests or LLM intent. These navigation capabilities are crucial for traversing application workflows and accessing different functionalities.
\end{itemize}

The design choice of employing an LLM to generate the action list stems from its ability to comprehend user intent expressed in natural language and plan a sequence of steps to achieve a desired goal \cite{Wang2023_conversational, Wen2024}. This approach contrasts with traditional methods that often require predefined tasks or extensive manual scripting \cite{Li2020_mapping}. By leveraging the LLM's reasoning capabilities and knowledge of common mobile UI patterns, the accessibility service can effectively execute complex, multi-step tasks without explicit low-level programming for every possible scenario \cite{Huang2023}. Furthermore, the LLM's ability to process the screen context, often represented as a simplified HTML structure \cite{Wang2023_conversational, Wen2024}, allows for context-aware action generation, ensuring that the service attempts only valid interactions with the currently displayed UI elements. This procedural execution of LLM-generated action lists enables a more flexible and intelligent approach to mobile task automation for accessibility purposes, offering a potential alternative or augmentation to traditional screen readers \cite{Ghosh2024}.

\subsection{Service Behavior}
\begin{figure}[h]
    \centering
    \includegraphics[width=0.5\linewidth]{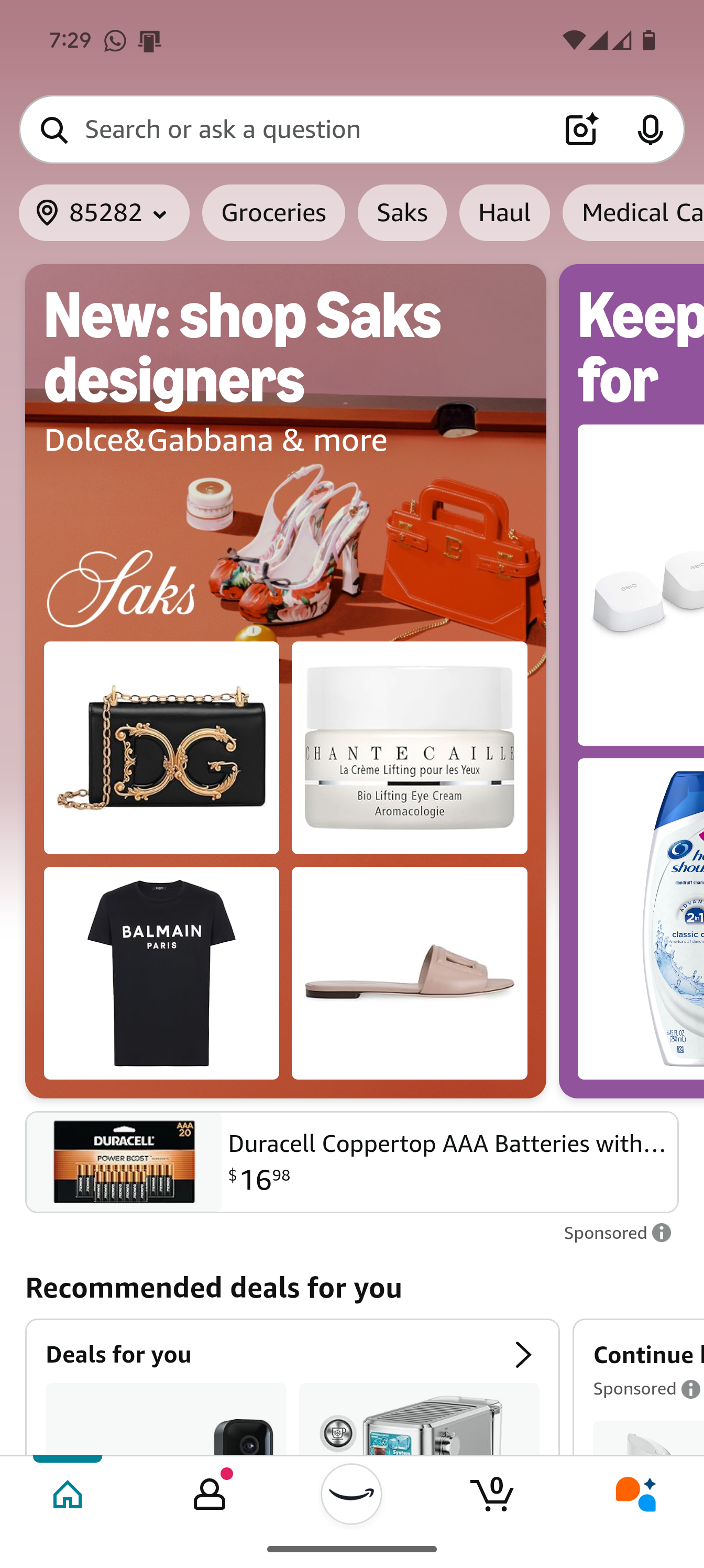}
    \caption{Amazon Shopping App \hfill \break
    Input: none (Default action summarize)}
    \label{fig:amazon}
\end{figure}
Response for the Amazon shopping query (Figure~\ref{fig:amazon}): 
\begin{quote}
\textbf{Insight}: The screen displays the Amazon.com homepage. At the top, there is a search bar labeled ``Search or ask a question'' with voice and scan options. Below, you'll find a sub-navigation bar with options such as ``Groceries'', ``Saks'', ``Haul'', ``Medical Care'', ``Same-Day'', ``Pharmacy'', ``In-Store Code'', ``Alexa Lists'', ``Prime'', ``Video'', and ``Music''. There are several sections with recommended deals and items, including ``New: shop Saks designers Dolce\&Gabbana \& more'', ``Keep shopping for'', ``Deals you'll love Based on your recent views'', ``Deals based on your lists'', ``Mother's Day is May 11 Chill mom gifts under \$50'', ``Continue shopping deals'', ``4+ star deals for you'', ``Summer Favorites Fashion finds under \$50'', ``Mother's Day is May 11 Top 100+ gifts'', ``More top picks for you'', ``Favorite Reordered Groceries'' and other deals, often with discount percentages mentioned. Also present is a sponsored ad for ``Duracell Coppertop AAA Batteries''. At the bottom is a tab navigation bar with the options Home, Your Amazon, Quick links, Cart, and Amazon Rufus.
\end{quote}

\begin{figure}[h]
    \centering
    \includegraphics[width=0.45\linewidth]{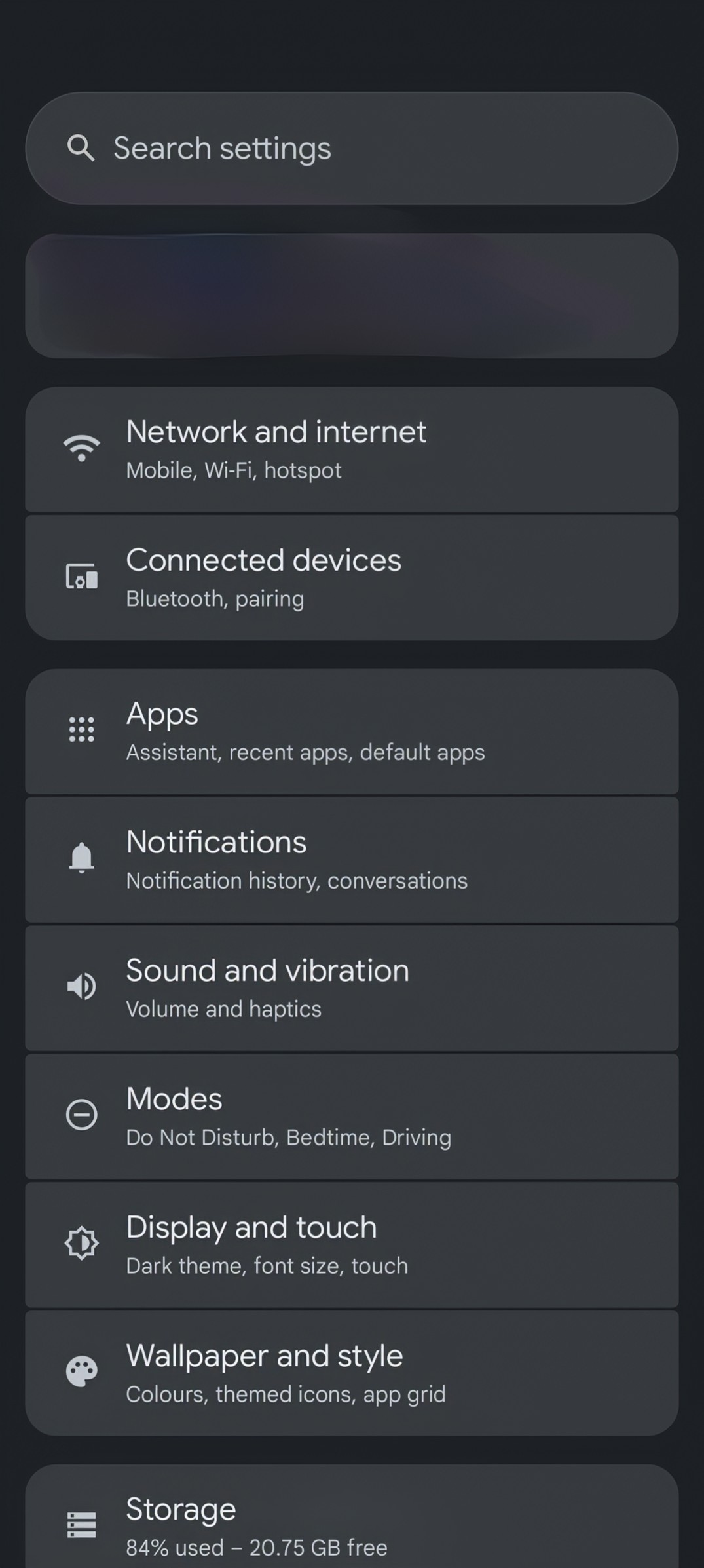}
    \hfill
    \includegraphics[width=0.45\linewidth]{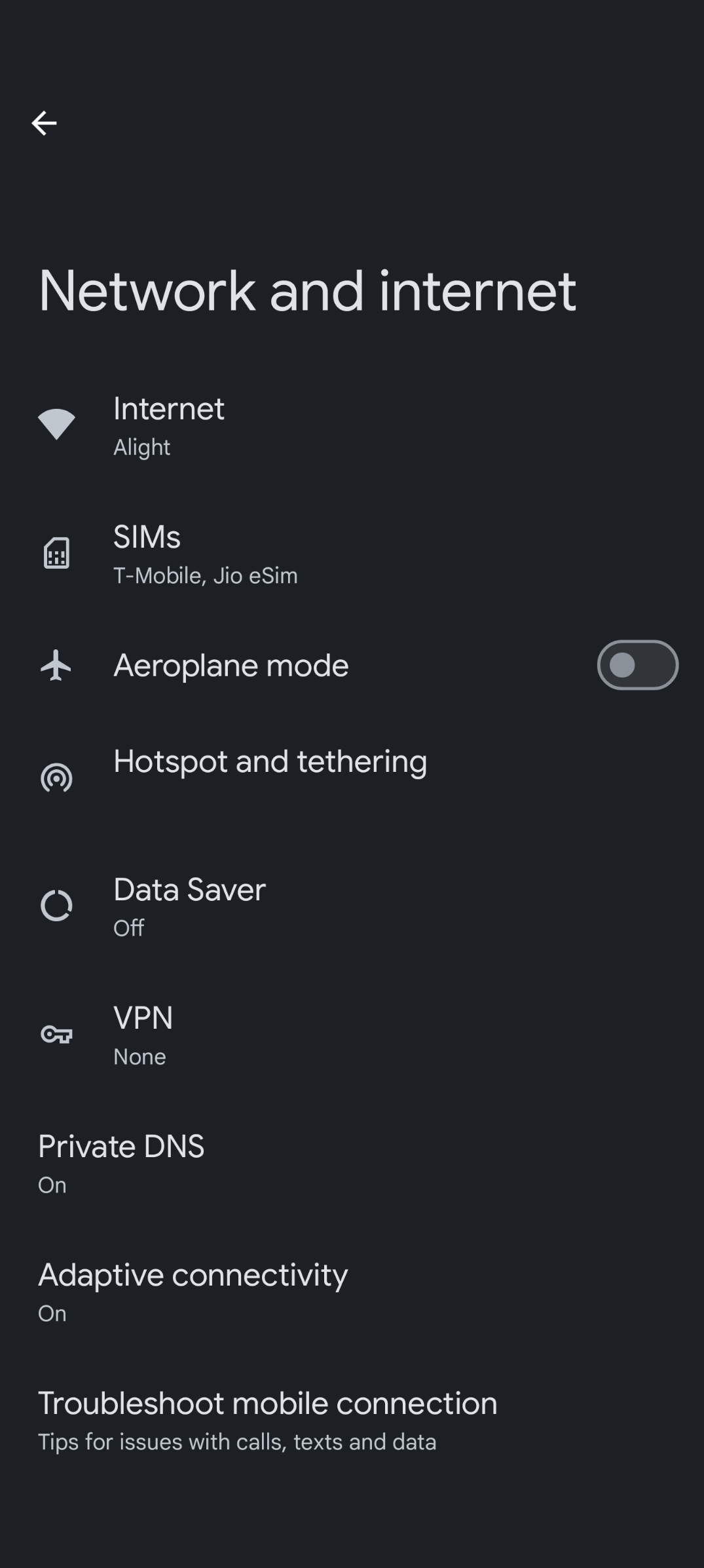}
    \caption{Settings App \hfill \break
    Input: Go to Network and Internet settings}
    \label{fig:settings_menu}
\end{figure}
Response for the settings query (Figure~\ref{fig:settings_menu}): 
\begin{quote}
    \textbf{Insight}: Okay, going to Network and internet.
    On the Network and internet screen, you can access the following options: Internet showing the network Alight, SIMs showing T-Mobile and Jio eSim, Hotspot and tethering, Data Saver which is off, VPN showing None, Private DNS which is on, Adaptive connectivity which is on, and Troubleshoot mobile connection, with tips for issues with calls, texts and data.
\end{quote}

\section{Methods}

\subsection{Participants}
The recruitment process involved sending a short Google form to members of the Human-Computer Interaction (HCI) Spring 2025 class at a South-Western University in the US via Slack, to express their interest and select convenient time slots to participate in the experiment. 11 participants were recruited for the study. The participants were between 20 and 30 years old, with 54.5\% being men and 45.5\% women. The experiments were carried out within two weeks, with each participant spending approximately one hour to complete both the experiment and the interview. Each participant received additional credit in the HCI course as compensation for their participation in the study.

\subsection{Experimental Setup}
A within-subject experimental design was used, allowing each participant to use both Talkback and our solution, Insight. Insight was installed on an android test device for each participant to interact with. Participants had to perform two pre-identified tasks, selected based on the design from a prior study called Savant, which has implemented LLMs for PC control via voice \cite{Ghosh2024}.  
\paragraph{\textit{\textbf{Task 1: Phone navigation and settings update}}}
Each participant had to navigate to the settings app on the test device and increase the media volume.

\paragraph{\textit{\textbf{Task 2: Shopping}}}
Each participant had to open the Amazon shopping application that had been pre-installed on the test device, and identify the item that had been added to the shopping cart.

Participants were asked to look away from the test device while completing each task, since they were not blind or visually impaired. Before the experiments began, the research team sought participants' consent to record the interactions and subsequent interview. Three team members were present for each user experiment: the first researcher explained how each accessibility service works to the participant and guided them through the experiment, the second researcher conducted the interviews after participants completed their tasks, and the third researcher was in charge of measuring the length of time it took the participant to complete each task.

\subsection{Interviews}
Each participant was asked a set of questions following the completion of both tasks. They were asked about their experience with using each service, such as the challenges they faced while completing the tasks and suggestions for the improvement of Insight. We also asked the participants which of the services was easier to use, more intuitive, and which service they would prefer for daily use. 

\subsection{Data Analysis}
The interviews were recorded using Zoom, and the transcripts were recovered from the transcription services offered by the platform. 

The research team analyzed the interview transcripts using inductive thematic analysis. Thematic analysis allowed for the identification of patterns and themes in the data \cite{Braun2006}, which was helpful in explaining the challenges participants faced while using Insight compared to Talkback.

The process involved two coders going over each interview transcript. The intercoder reliability, kappa, was calculated to be 0.64, which was a satisfactory level of agreement between the two coders. Coding was done in Google Docs, using In vivo and process coding techniques. The two coders initially developed 51 codes, such as ``tedious swiping,'' ``tough navigation,'' ``praising speed,'' and ``talking directly.'' All codes were further categorized into major themes under advantages and limitations of each service. For example, codes like ``tedious swiping,'' ``frustrating to click,'' and ``Scrolling was challenging'' were categorized under ``difficult gesture navigation.'' Also, the codes ``voice was better,'' ``familiar like Siri,'' and ``talking to a companion'' are categorized under ``Convenient to interact with.''

\section{Results}

\subsection{Task Efficiency: Completion Time}

\begin{figure}
    \centering
    \includegraphics[width=1\linewidth]{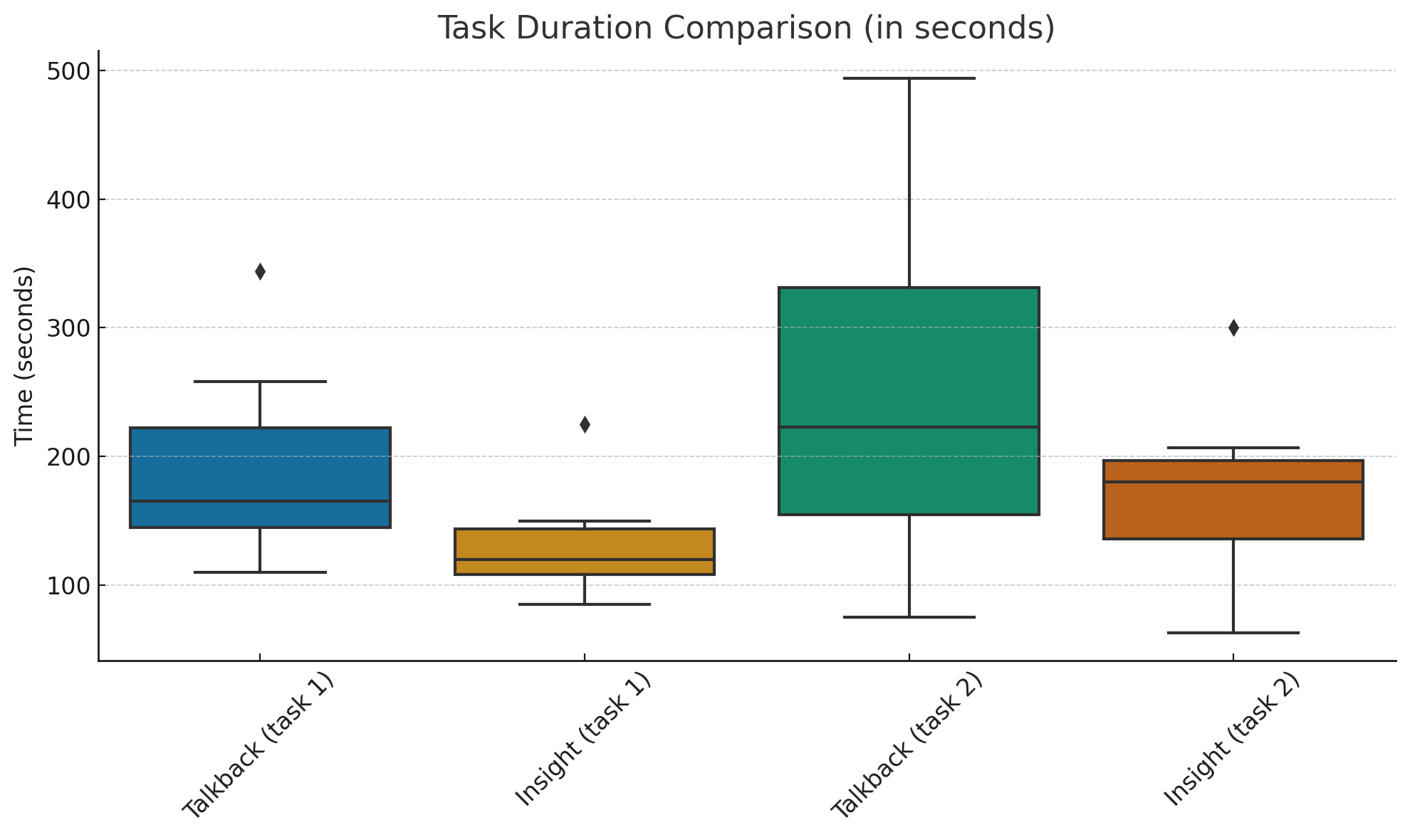}
    \caption{Task Duration}
    \label{fig:task_duration}
\end{figure}

To evaluate the efficiency of Insight relative to TalkBack, we recorded the time taken (in seconds) to complete task 1 and task 2. A boxplot visualization (\textbf{Figure~\ref{fig:task_duration}}) highlights a consistent trend: participants completed tasks more quickly using Insight than with TalkBack.

For example, in Task 2 (Online Shopping), TalkBack exhibited high variance, with completion times ranging from 75 to nearly 500 seconds. In contrast, Insight enabled users to complete the same task in substantially less time and with narrower variance, suggesting lower cognitive overhead and faster interaction cycles. This performance improvement can be attributed to Insight's screen summarization and natural language input, which eliminate the need for sequential scanning of UI elements.

\subsection{Perceived Usability and Adoption Potential}

Following task completion, participants rated both services across three dimensions: \textit{ease of use}, \textit{intuitiveness}, and \textit{likelihood of daily use}. A grouped bar chart (\textbf{Figure~\ref{fig:ratings}}) summarizes these ratings as percentages:

\begin{figure}
    \centering
    \includegraphics[width=1\linewidth]{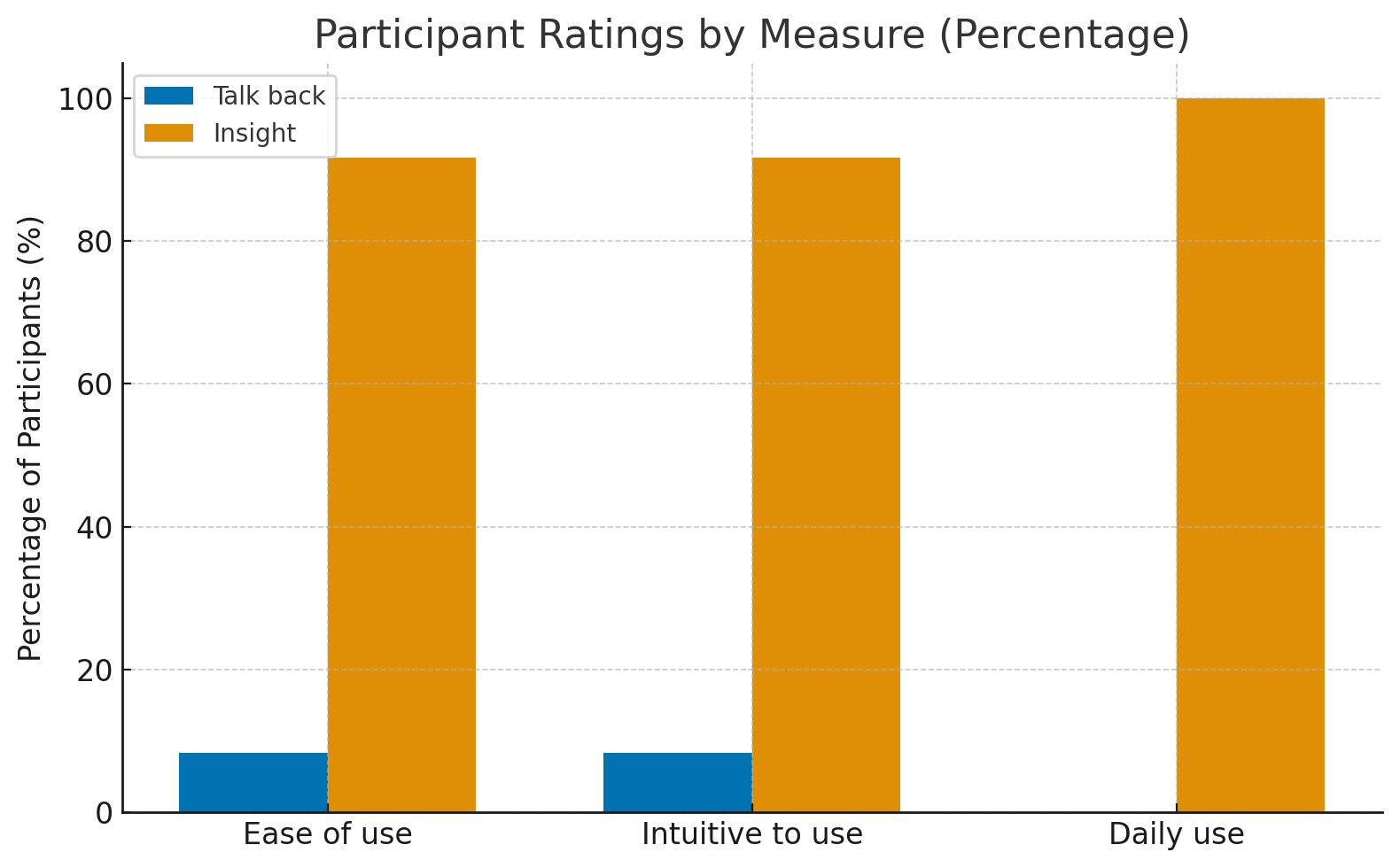}
    \caption{Participant Rating by Measure}
    \label{fig:ratings}
\end{figure}

\begin{itemize}
    \item \textbf{Ease of Use:} Only 8\% of participants found TalkBack easy to use, compared to 92\% for Insight.
    \item \textbf{Intuitiveness:} TalkBack was rated intuitive by 8\% of participants, whereas Insight scored 92\%.
    \item \textbf{Daily Use Likelihood:} 100\% of participants expressed willingness to use Insight daily, while none preferred TalkBack for regular use.
\end{itemize}

These results indicate that Insight offers a more accessible and natural interaction model than TalkBack. Participants attributed this to Insight's conversational dialogue interface and real-time summarization capabilities.

\subsection{Qualitative Observations}

To complement the quantitative findings, we conducted semi-structured interviews with participants to better understand their lived experience using both accessibility services. Four thematic categories emerged: limitations and advantages of TalkBack, and limitations and advantages of Insight.

\subsubsection{Limitations of TalkBack}

Participants expressed significant frustration with TalkBack's sequential navigation model, which required numerous swipes and made it cognitively demanding to locate UI elements. One participant described it as:

\begin{quote}
``I had to swipe a lot, and it was confusing\ldots what I wanted to do didn't happen until much later'' (Participant 1).
\end{quote}

Another noted that the gesture system often misfired or caused disorientation:

\begin{quote}
``TalkBack\ldots was going to everything sequentially. Sometimes our swipes may be misplaced\ldots I was not able to understand where it is going exactly'' (Participant 2).
\end{quote}

Users also emphasized the high cognitive load and time consumption associated with scanning every screen element:

\begin{quote}
``If a person is blind\ldots scrolling left, left, left or right, right, right\ldots it's very frustrating\ldots a one-minute task can take 20 to 30 minutes'' (Participant 3).
\end{quote}

\subsubsection{Advantages of TalkBack}

Despite these challenges, a few participants found that TalkBack's design to announce only highlighted content could sometimes reduce information overload:

\begin{quote}
``TalkBack was better for settings\ldots it was very simple. It was done just in an instant'' (Participant 2).
\end{quote}

This control over auditory focus was seen as useful in scenarios requiring precision or reduced distraction.

\subsubsection{Limitations of Insight}

Although Insight was generally preferred, participants reported usability issues related to interactivity and speech comprehension. A common concern was the inability to interrupt the system during lengthy screen summarizations:

\begin{quote}
``I don't have the option to interrupt\ldots it starts summarizing again and again'' (Participant 1).
\end{quote}

Some users also struggled with command recognition, noting that specific phrasing was required for successful interaction:

\begin{quote}
``You need to give the right command\ldots simpler, more general conversation is needed'' (Participant 2).
\end{quote}

Privacy concerns also emerged due to the application's use of cloud-based large language models:

\begin{quote}
``You should give a consent\ldots some people might be more concerned about privacy because of their buying details and everything'' (Participant 3).
\end{quote}

\subsubsection{Advantages of Insight}

Participants widely praised Insight for its natural voice interaction and effective summarization. Users emphasized its ease of use and suitability for everyday tasks:

\begin{quote}
``It was convenient\ldots just saying what to do and it responds. It's better than remembering gestures'' (Participant 4).
\end{quote}

Insight also significantly reduced cognitive load by offering high-level summaries of the screen content:

\begin{quote}
``It just gives a brief overview\ldots TalkBack was not summarizing anything. You had to go sequentially'' (Participant 2).
\end{quote}

\begin{quote}
``Your app was more natural\ldots even if the person doesn't know how to use a phone, it would be helpful for them'' (Participant 3).
\end{quote}

These findings illustrate that Insight provides a more intuitive and inclusive user experience, but with opportunities to improve flexibility, privacy, and conversational robustness.

\section{Discussion}
The study points to four clear lessons for the next generation of mobile screen-access tools and shows how our findings build on, extend, or differ from earlier research.

\subsection{Efficiency and User Preference}
\textit{Insight} cut task-completion time by 26--29\% and was chosen in more than 90\% of trials.  
This mirrors desktop gains reported for \textbf{Savant}---where blind users reached controls almost four-times faster and scored 3$\times$ higher on the Single Ease Question (SEQ)~\cite{Ghosh2024}---and mobile gains for \textbf{GPTVoiceTasker}, which lifted real-world task efficiency by 34.9\%~\cite{Vu2024}.  
Together, these results underline that conversational, LLM-powered interfaces reliably speed up work across devices.

\subsection{Long-Term Familiarity and Adaptation}
LLMs give rapid ``out-of-the-box'' summaries, but several studies stress the need for adaptive verbosity as users become experts.  
Savant's authors note that advanced participants wanted briefer answers over time, and \textbf{GPTVoiceTasker} lets people edit or shorten saved commands~\cite{Vu2024}.  
Our participants shared similar opinions, so future versions of Insight should learn usage patterns and adjust its wording and pace automatically.

\subsection{Cognitive Load Reduction}
Allowing people to say \emph{what} they want instead of \emph{where} to tap shifts the heavy thinking from user to model.  
Lower \textbf{NASA-TLX} workload scores for Savant, major reductions in perceived effort for \textbf{GPTVoiceTasker}, and simpler task breakdowns in systems like \textbf{AutoDroid} \cite{Wen2024} all support this claim.  
Our study adds fresh evidence on native apps, showing that intent-level commands indeed ease the mental burden on small touch screens.

\subsection{Privacy through Hybrid Navigation}
Participants liked the idea of mixing TalkBack's element-by-element focus (quiet in public) with Insight's full-screen summaries (rich context when privacy allows).  
Web-based \textbf{PROTECT} also found that users sometimes toggled between brief structural hints and detailed chat guidance~\cite{Costabile2024}.  
This suggests that a \emph{mode-switch}---gesture for pinpoint output, conversation for overview---can offer both discretion and power.

\subsection{Limitations}
\begin{itemize}
    \item \textbf{Participant sample.} All eleven participants were sighted volunteers who were asked not to look at the test device; their feedback lacks the lived expertise of BVI users.
    \item \textbf{Platform scope.} Insight currently runs only on Android; cross-platform performance remains untested.
    \item \textbf{Environmental control.} Due to the location of the study, the campus library, background noise occasionally degraded voice recognition, potentially inflating error rates.
    \item \textbf{Study duration.} Short-term exposure cannot reveal longitudinal issues such as fatigue or evolving trust.
    \item \textbf{Network reliance.} The service streams screen content to a cloud-hosted model. Hence the privacy and latency may differ in real-world deployment.
\end{itemize}

\subsection{Future Work}
To translate Insight's promise into a deployable, inclusive service, we outline seven concrete research and engineering directions.

\begin{enumerate}
    \item \textbf{Hybrid Gesture--Conversational Interaction Model.}
    Guided by participants' enthusiasm for a ``tap-then-ask'' workflow, future design can let users 
    \begin{itemize}
        \item Tap once to set the TalkBack focus,
        \item Request a targeted or global LLM summary, and
        \item Issue an intent-level command---all within the same interaction loop.  
    \end{itemize}
    The study can compare this hybrid model with Insight-only and TalkBack-only baselines on speed, cognitive load, and perceived privacy.

    \item \textbf{Interactivity and Interruptions.}
    Participants asked to \emph{pause}, \emph{skip}, or \emph{fast-forward} long responses. A lightweight protocol that recognizes brief voice cues (e.g., ``stop'', ``next'') can be added to the service to interrupt the screen summarization by the user as soon as the user has determined the next action/question. This way the user does not have to listen to the entire response. This would help reduce the time to perform tasks for the users.

    \item \textbf{Privacy Evaluation.}
    The present study does not audit what happens to the screen captures and speech transcripts that Insight transmits to its remote LLM back-end; thus we cannot yet assure users that no personally sensitive information is stored, re-used for model training, or leaked. 
    Future iterations can conduct a formal privacy-impact assessment that (i) maps every data flow end-to-end and (ii) establishes a data-retention policy aligned with GDPR and CCPA principles of data minimization.

    \item \textbf{On-device LLM Deployment.}
    A local model removes network dependence, reduces latency and enables offline usage (e.g., during flights). Models in beta versions such as Gemma-3 could be utilized for the LLM engine in the service.

    \item \textbf{Comparative Model Benchmarking.}
    Insight's pipeline can potentially support different LLM engines (Gemma-3, Llama 3, GPT-4o). A study to benchmark the performance with different LLMs could be conducted.

    \item \textbf{Context-aware Summaries.}
    By fusing mobile app specific information and data (GPS, activity recognition, usage logs) with conversation history, Insight could produce shorter, situation-specific prompts---e.g., emphasizing transit data at bus stops or highlighting unread messages during work hours.

    \item \textbf{Longitudinal BVI Field Study.}
    Recruitment of a diverse cohort of blind and low-vision users for a multi-week long deployment on both Android and iOS, collecting interaction logs, experience-sampling data, and interviews to uncover learning curves, trust dynamics, and breakdowns.
\end{enumerate}

These avenues aim to stabilize Insight, tighten its privacy guarantees, broaden device coverage, and ground its evolution in the lived realities of BVI users.

\section{Conclusion}

The study demonstrated that LLM-driven accessibility services like Insight significantly enhance mobile usability among visually impaired and blind users. Users completed tasks more efficiently and with reduced cognitive effort while using Insight compared to default aids like TalkBack. Quantitative data revealed reduced task execution times and greater user preference in terms of usability aspects. Qualitatively, users liked the dialogue-and-intent-based interaction model and summarization of the screen in real time, but felt the need for interruptibility and showcased concerns for privacy about sensitive screen data.

These findings suggest that LLMs can realistically transform the mobile accessibility model from a gesture-heavy sequential interaction to a high-level seamless interaction. To be successful, however, it will rely on design accommodations for long-term user familiarity, voice and gesture privacy protection, and hybrid models of interaction that couple voice with gestures.

Finally, Insight shows an effective way forward for accessible mobile computing, both by extending technical interaction and providing BVI users with more natural, independent, and human-based control over their devices. Future work should explore long-term use by real BVI users and adjust the system's responsiveness, privacy features, and flexibility.

\end{document}